\documentclass[11pt]{article}
\usepackage{a4}
\usepackage{mathrsfs}

\newcommand{\nit}{\noindent}
\newcommand{\nl}{\newline}
\newcommand{\np}{\newpage}
\newcommand{\dsp}{\displaystyle}
\newcommand{\vs}[1]{\vspace{#1 ex}}
\newcommand{\hs}[1]{\hspace{#1 em}}
\newcommand{\bfr}{\begin{flushright}}
\newcommand{\efr}{\end{flushright}}
\newcommand{\bc}{\begin{center}}
\newcommand{\ec}{\end{center}}
\newcommand{\ben}{\begin{enumerate}}
\newcommand{\een}{\end{enumerate}}

\newcommand{\be}{\begin{equation}}
\newcommand{\ee}{\end{equation}}
\newcommand{\ba}{\begin{array}}
\newcommand{\ea}{\end{array}}
\newcommand{\ct}{\cite}
\newcommand{\bit}{\bibitem}

\newcommand{\ag}{\alpha}
\newcommand{\bg}{\beta}
\newcommand{\gam}{\gamma}

\newcommand{\ve}{\varepsilon}
\newcommand{\zg}{\zeta}
\newcommand{\thg}{\theta}
\newcommand{\kg}{\kappa}
\newcommand{\lb}{\lambda}
\newcommand{\sg}{\sigma}
\newcommand{\rg}{\rho}

\newcommand{\og}{\omega}
\newcommand{\Gam}{\Gamma}
\newcommand{\Del}{\Delta}

\newcommand{\bfxi}{\mbox{{\boldmath $\xi$}}}

\newcommand{\lh}{\left(}
\newcommand{\rh}{\right)}

\begin{document}

\pagestyle{empty}
\begin{flushright}
NIKHEF/02-008
\end{flushright} 
\vs{2} 

\begin{center}
{\Large{\bf{Particle motion}}} \\ 
\vs{3} 
{\Large{\bf{in electro-magnetic and gravitational pp-waves}}}\\
\vs{10}

{\large M.\ Faquir$^{*}$ and J.W.\ van Holten$^{\dagger}$} \\
\vs{2}

{\large{NIKHEF, Amsterdam NL}} \\
\vs{2}

\today 
\vs{10}

{\small{ \bf{Abstract} }} \\
\end{center}

\nit
{\footnotesize{We discuss the motion of neutral and charged particles in a plane 
    electro-magnetic wave and its accompanying gravitational field. }}
\vfill
\footnoterule 
\nit
{\footnotesize{
$^{*}$ \tt{e-mail: mohamedf@nikhef.nl} \\
$^{\dagger}$ \tt{e-mail: v.holten@nikhef.nl}}}

\np
~\hfill 

\np
\pagestyle{plain} 
\pagenumbering{arabic}

\section{Introduction} 

\nit
The existence of plane-fronted parallel wave solutions in general
relativity has been known since a long time \ct{brinkmann}-\ct{ksmch}. 
These solutions are not only interesting by themselves, but also 
because similar solutions accompany massless fields of lower spin,
such as electro-magnetic, scalar and spinor fields \ct{ksmch},
\ct{jwvh1}-\ct{jwvh3}. 

Gravitational $pp$-waves can be described by space-time metrics  
   \be  
   g_{\mu\nu} dx^{\mu} dx^{\nu} = -\, du dv - K(u,x,y) du^2 +  
   dx^2 + dy^2 ,  
   \label{1}  
   \ee 
where the light-cone co-ordinates $(u,v)$, transverse to the $x$-$y$-plane, 
are related to time- and longitudinal co-ordinates $(t,z)$ by:  
\be  
u = ct - z, \hspace{3em} v = ct + z.  
\label{2}  
\ee    
There are similar solutions with the roles of $v$ and $u$ interchanged, 
which we will not discuss explicitly here. 
With the metric (\ref{1}), the only non-vanishing connection co-efficients are:  
\be  
\Gam_{uu}^{\;\;\;\;v} = K_{,u}, \hspace{2em}  
\Gam_{uu}^{\;\;\;\;x} = \frac{1}{2}\, \Gam_{xu}^{\;\;\;\;v} =  
\frac{1}{2}\, K_{,x}, \hspace{2em}  
\Gam_{uu}^{\;\;\;\;y} = \frac{1}{2}\, \Gam_{yu}^{\;\;\;\;v} =  
\frac{1}{2}\, K_{,y}.  
\label{3}  
\ee  
The free gravitational $pp$-waves are solutions of the free Einstein equations, 
which for the metrics (\ref{1})
simplify to 
\be
\Del_{trans} K = K_{,xx} + K_{,yy} = 0.
\label{4}
\ee
The simplest non-trivial one is 
\be 
K(u,x,y) = \frac{1}{2}\, \kg_+(u) (x^2 - y^2) + \kg_{\times}(u)\, xy .
\label{5}
\ee 
Here $\kg_{(+,\times)}(u)$ are the amplitudes of the two different polarization
states, as appropriate to quadrupole-type waves propagating at the speed of light. 

Metrics of the type (\ref{1}) are also found in the solution of the coupled
Maxwell-Einstein, massless Klein-Gordon-Einstein and massless Dirac-Einstein 
equations. As a physically interesting example, consider the gravitational field 
of a light wave propagating along the $z$-axis. The light-wave is characterized 
completely by a vector potential 1-form
\be 
{\bf A} = \sin ku\, (a_x dx + a_y dy),
\label{6}
\ee 
with $(a_x, a_y)$ the components of the constant transverse polarization vector.
The corresponding electric and magnetic fields are 
\be 
{\bf E}(u) = \og {\bf a}\, \cos ku, \hs{2} 
{\bf B}(u) = {\bf k} \times {\bf a}\, \cos ku, 
\label{7}
\ee 
where $\og = kc$. These fields provide a solution of the coupled Maxwell-Einstein with 
a metric of the form (\ref{1}), with  
\be 
\ba{lll}
K(u,x,y) & = & \dsp{ \frac{2\pi \ve_0 G}{c^4}\,  \lh {\bf E}^2(u) + c^2 {\bf B}^2(u) 
 \rh\, \lh x^2 + y^2 \rh  }\\
 & & \\
 & = & \dsp{ 
 \frac{2\pi \ve_0 G}{c^2}\, k^2 {\bf a}^2\, (1 + \cos 2ku)\, \lh x^2 + y^2 \rh. }
\ea 
\label{8}
\ee 
The motion of a particle of mass $m$ and charge $q$ propagating in gravitational and
electro-magnetic background fields is governed, in the limit of neglecting radiation
reaction terms, by the covariant Lorentz force
\be 
\ddot{x}^{\mu} + \Gam_{\lb\nu}^{\;\;\;\;\mu} \dot{x}^{\lb} \dot{x}^{\nu} 
 = \frac{q}{m}\, F^{\mu}_{\;\;\nu} \dot{x}^{\nu}. 
\label{9}
\ee 
Here the overdot denotes a derivative w.r.t.\ proper time, defined
in the usual way from the universal conserved value of the total four-velocity:
\be 
u_{\mu}^2 = g_{\mu\nu} \frac{dx^{\mu}}{d\tau}\, \frac{dx^{\mu}}{d\tau}\, = - c^2.
\label{1.0}
\ee 
We apply these equations to study the motion of a particle in the background of the
waves fields (\ref{1},\ref{6},\ref{7}). Recalling the results of ref.\ct{jwvh3},
the light-cone co-ordinate $u$ can be used as a proper-time variable via the relation  
\be 
\dot{u} = \gam = \mbox{const.} \hs{1} \Rightarrow \hs{1} u = \gam \tau. 
\label{1.1}
\ee 
The expression (\ref{1.0}) for the total four-velocity then leads to the following
relation between proper time and laboratory time:
\be 
\frac{dt}{d\tau} = \gam \frac{dt}{du} = \sqrt{\frac{1 - \gam^2 K/c^2}
{1 - {\bf v}^2/c^2}},
\label{1.2}
\ee 
showing that there is both a gravitational and a kinematic time dilation. Substitution
back into eq.(\ref{1.0}) allows one to cast it into a true conservation law of the form
\be 
h \equiv K + \frac{1 - {\bf v}^2/c^2}{(1 - v_z/c)^2} = \frac{c^2}{\gam^2}. 
\label{1.3}
\ee 
Two independent equations of motion remain, which we take to be those for the tranverse 
co-ordinates ${\bfxi} = (x,y)$. The Lorentz-Einstein equation (\ref{9}) for these
components becomes 
\be 
\frac{d^2\bfxi}{du^2}\, = - \frac{2\pi \ve_0 G}{c^2}\, k^2 {\bf a}^2 
 \lh 1 + \cos 2ku \rh \bfxi - \frac{q}{m\gam} k {\bf a} \cos ku. 
\label{1.4}
\ee
We can define the components w.r.t.\ the polarization 
of the light wave as 
\be 
\bfxi = \bfxi_{\|} + \bfxi_{\perp}, \hs{2} 
\bfxi_{\|} = \frac{\bfxi \cdot {\bf a}}{|{\bf a}|^2}\, {\bf a}, \hs{2} 
\bfxi_{\perp} = \frac{\bfxi \times {\bf a}}{|{\bf a}|}. 
\label{1.5}
\ee 
Under this decomposition eq.(\ref{1.4}) splits into a homogeneous and an inhomogeneous 
equation 
\be 
\ba{l}
\dsp{ \frac{d^2\bfxi_{\perp}}{du^2} + \frac{2\pi \ve_0 G}{c^2}\, k^2 {\bf a}^2 
 \lh 1 + \cos 2ku \rh \bfxi_{\perp} = 0, }\\
  \\
\dsp{ \frac{d^2\bfxi_{\|}}{du^2} + \frac{2\pi \ve_0 G}{c^2}\, k^2 {\bf a}^2 
 \lh 1 + \cos 2ku \rh \bfxi_{\|} = - \frac{q}{m\gam} k {\bf a} \cos ku. }
\ea
\label{1.6}
\ee 
In refs.\ \ct{jwvh2,jwvh3} it was observed, that in the static limit $k \rightarrow 0$
taken such that $E_0 = c B_0 = ck |{\bf a}| =$ constant, one obtains a simple harmonic 
motion with frequency $\nu$ (in Hz) given by: 
\be 
2 \pi \nu = \sqrt{\frac{4\pi \ve_0 G}{c^2}}\, E_0 = 0.3 \times 10^{-18} E_0\, 
 \mbox{(V/m)}. 
\label{1.7}
\ee 
In this paper we discuss the motion in the non-static case $k \neq 0$.

\section{Transverse motion}

We first focus on the case $q=0$, describing a neutral massive test 
particle in an  electro-magnetic and gravitational $pp$-wave. According to 
(\ref{1.6}), the transverse coordinates $(x, y)$  all obey the same homogeneous 
Mathieu equation. Besides, we choose to restrict our study to the case where 
the only contribution to the metric co-efficient $K$ is the electromagnetic 
one, given by (\ref{8}). In other words, no free gravitational $pp$-waves are 
taken to be present: $\kg_{+,\times}(u)$=0. It then follows that the 
motion of the particle is invariant under the exchange of the tranverse 
coordinates and rotations in the transverse plane. The equation for 
$\xi= (x, y)$ now reads (with $a^2 = {\bf a}^2$):
\be
\dsp{ \frac{d^2\xi}{du^2} + \frac{2\pi \ve_0 G}{c^2}\, k^2 a^2 
 \lh 1 + \cos 2ku \rh \xi = 0, }\\
\label{1.8}
\ee
A solution procedure for eq.(\ref{1.8}) in the weak-coupling regime was presented in 
ref.\ \ct{jwvh2}. Here we discuss the construction of the full solution in terms of a 
generalized Fourier series expansion. Because of the particular form of the equation 
(\ref{1.8}), we can construct a series in terms of either even or odd multiples of the 
fundamental wave factor $k$. In this section we consider the generalized odd solutions, 
which can be parametrized as 
\be
\dsp{ \xi(u) = e^{\sg k u} \sum_{n=0}^{\infty}\, \lh \xi_n \cos (2n+1) ku + \eta_n 
 \sin (2n+1) ku \rh,}
\label{2.1}
\ee
with real Fourier components $(\xi_n, \eta_n)$.  Note that for $\sg^2 \leq 0$ (and 
taking the real part) one obtains solutions which are purely of trigonometric type; 
for $\sg^2 > 0$ the solutions describe oscillations enhanced by parametric resonance.  
The other type of solutions having even Fourier part is discussed in the appendix. 
The substitution (\ref{2.1}) converts the Mathieu equation into an infinite-dimensional 
linear matrix equation
\be 
\lh \ba{cccc} A_0 &  B  &  0  & ... \\
               B  & A_1 &  B  & ... \\
               0  &  B  & A_2 & ... \\
               .  &  .  &  .  &  .  \ea \rh \lh \ba{c} \rg_0 \\ 
                                                       \rg_1 \\
                                                       \rg_2 \\ 
                                                        . \ea \rh\, = 0, 
\label{2.2}
\ee 
where the blocks $A_m$ and $B$ are $2 \times 2$ matrices given by 
\be 
\ba{rcl} 
A_0 & =  & \dsp{ \lh \ba{cc} \sg^2 - 1 + \frac{3\kg^2}{4} & 2\sg  \\
 -2 \sg & \sg^2 -1 + \frac{\kg^2}{4} \ea\rh, }\\
 & & \\
A_m & = & \dsp{ \lh \ba{cc} \sg^2 - (2m + 1)^2 + \frac{\kg^2}{2} & 2 (2m+1) \sg  \\
 -2 (2m + 1) \sg & \sg^2 -(2m+1)^2 + \frac{\kg^2}{2} \ea\rh, \hs{1} (m \geq 1), }
\ea 
\label{2.3}
\ee 
and 
\be 
B = \frac{\kg^2}{4}\, \lh \ba{cc} 1 & 0 \\ 
                                  0 & 1 \ea \rh, \hs{2} 
\rg_m = \lh \ba{c} \xi_m \\ \eta_m \ea \rh. 
\label{2.3.1}
\ee 
A solution of the system (\ref{2.2}) exists provided the linear matrix operator 
has zero-modes. To determine their existence we cast the equation into
upper-triangular form: 
\be 
\lh \ba{cccc} a_0 &  B  &  0  & ... \\
               0  & a_1 &  B  & ... \\
               0  &  0  & a_2 & ... \\
               .  &  .  &  .  &  .   \ea \rh\, \lh \ba{c} \bar{\rg}_0 \\
                                                          \bar{\rg}_1 \\
                                                          \bar{\rg}_2 \\
                                                             . \\ \ea \rh\, = 0.
\label{2.3.2}
\ee 
The $2 \times 2$ matrices $a_m$ can be expressed in terms of continuing fractions as
\be 
a_m = A_m - B^2 \left[ A_{m+1} - B^2 \left[ A_{m+2} - ... \right]^{-1} \right]^{-1}, 
\label{2.3.3}
\ee 
and the basis of vectors has been transformed by $\bar{\rg}_0 = \rg_0$, and
\be 
\bar{\rg}_m = \rg_m + a^{-1}_{m} B \rg_{m-1}, \hs{1} (m \geq 1).  
\label{2.3.4}
\ee 
This transformation is singular if and only if any of the $a_m$ $(m \geq 1)$ has 
zero-modes, which is precisely the condition for the Mathieu equation to have 
non-trivial solutions. 

Zero-modes of $a_0$ represent the lowest-frequency solutions of the Mathieu equation; 
they exhibit parametric resonance for $\sg > 0$. The existence of zer-modes implies
that the determinant of $a_0$ vanishes. This determinant can be computed in successive 
approximations by truncating the infinite fraction (\ref{2.3.3}). The zeroth-order 
approximation is
\be 
a_0 = A_0 \hs{1} \Rightarrow \hs{1} 
\left| a_0^{(0)} \right| = (\sg^2 + 1)^2 + \kg^2 (\sg^2 - 1) + \frac{3\kg^4}{16}. 
\label{2.4.1}
\ee 
The first- and higher-order approximations are of the form
\be 
a_0 = \lh \ba{cc} \sg^2 - 1 + \frac{3\kg^2}{4} + \kg^4 \Del _{11} & 2 \sg + \kg^4 
 \Del_{12} \\                 
 - (2\sg + \kg^4 \Del_{12}) & \sg^2 - 1 + \frac{\kg^2}{4} + \kg^4 \Del_{11}
 \ea \rh + {\cal O}(\kg^6), 
\label{2.4.0}
\ee 
where the quantities $\Del_{ij}(\sg)$ are rational functions of $\sg^2$;  
the determinant then is
\be 
|a_0| = (\sg^2 + 1)^2 + \kg^2 (\sg^2 - 1) + \kg^4 \lh \frac{3}{16}
 - \frac{\Del}{8}  \rh + {\cal O}(\kg^6).
\label{2.4.0.1}
\ee 
In first order  $a^{(1)}_0 = A_0 - B^2 A_1^{-1}$, with  
\be
\Del^{(1)}_{11} = \frac{1}{48}\, \frac{9 - \sg^2}{27 + 6 \sg^2 + \sg^4/3}, \hs{2}
\Del^{(1)}_{12} = \frac{1}{8}\, \frac{1}{27 + 6 \sg^2 + \sg^4/3},
\label{2.4.0.0}
\ee 
and 
\be 
\Del^{(1)} = \frac{\sg^4 - 22\sg^2 + 9}{\sg^4 + 18 \sg^2 + 81}. 
\label{2.4.2}
\ee 
Similarly, in second order with
\be
a_0 = \lh A_0 A_1 A_2 - B^2(A_0 + A_2) \rh \lh A_1 A_2 - B^2 \rh^{-1},  
\label{2.4.0.2}
\ee
one finds up to terms ${\cal O}(\kg^6)$:
\be 
\ba{rcl} 
 - 16 (\sg^2  - 15)\, \Del^{(2)}_{11} & = & \dsp{ 
 \frac{5}{8}\, + \frac{3}{8}\, \frac{(\sg^2 + 25)(\sg^4 - 94 \sg^2 + 225)}
 {(\sg^4 - 94 \sg^2 + 225)^2 + 256 \sg^2 (\sg^2 - 15)^2}, }\\
  & & \\
\Del^{(2)}_{12} & = & \dsp{ \frac{3}{8}\, 
 \frac{(\sg^2 + 25)^2}{(\sg^4 - 94 \sg^2 + 225)^2 + 256 \sg^2 (\sg^2 - 15)^2}. } 
\ea 
\label{2.4.2.0}
\ee 
Then in the determinant 
\be 
\Del^{(2)} = 
 \frac{(\sg^2 - 1) [(\sg^2 - 25)(\sg^4 - 94 \sg^2 + 225) + 160 \sg^2 (\sg^2 -15)]}{
  (\sg^4 - 94 \sg^2 + 225)^2 + 256 \sg^2 (\sg^2 -15)^2}.
\label{2.4.3}
\ee 
Observe that in both first and second order approximation
\be 
\lim_{\sg^2 \rightarrow 0} \Del^{(k)} = \frac{1}{9}, \hs{2} 
\lim_{\sg^2 \rightarrow \infty} \Del^{(k)} = 1, \hs{2} k = (1,2).
\label{2.4.4}
\ee 
From the above analysis it follows, that up to terms of order $\kg^6$ a solution 
for $\sg^2$ exists for which the determinant vanishes: $|a_0| = 0$, as given by 
the expression (\ref{2.4.0.1}). This condition has approximate solutions 
\be
\sg^2 = - \lh 1 + \frac{\kg^2}{2} \rh \pm\,
 \sqrt{2\kg^2 \lh 1 + \frac{\kg^2}{32}\, (1 + 2 \Del) \rh}.
\label{2.5.1}
\ee 
Positive non-zero values of $\sg^2$ exist only for the positive square root, provided 
\be 
2\kg^2 \lh 1 + \frac{\kg^2}{32}\, (1 + 2 \Del) \rh > \lh 1 + \frac{\kg^2}{2} \rh^2.
\label{2.5.1.1}
\ee 
or
\be 
\lh 1 - \frac{\kg^2}{2}\,  + \frac{\kg^2}{4}\, \sqrt{1 + 2\Del} \rh 
 \lh 1 - \frac{\kg^2}{2}\, - \frac{\kg^2}{4}\,  \sqrt{1 + 2\Del} \rh < 0. 
\label{2.5.2}
\ee 
Hence $\kg^2$ must be in the range 
\be 
\frac{1}{2 + \sqrt{1 + 2 \Del}}\, < \frac{\kg^2}{4}\, < \frac{1}{2 - \sqrt{1 + 2\Del}}. 
\label{2.5.3}
\ee 
Purely periodic solutions of the Mathieu equation exist for  a discrete set of values 
of $\kg$ for which $\sg = 0$; for these values the inequality (\ref{2.5.1.1}) turns 
into and equality. For all other values outside the range (\ref{2.5.3}) the solutions 
for $\sg$ are imaginary; the series (\ref{2.1}) then is purely trigonometric, but for 
non-rational $\sg$ the result is not periodic.  

From the inequality (\ref{2.5.3}) it follows, that the smallest possible value of 
$\kg^2$ for which parametric resonance occurs is of the order unity: $\kg^2 \geq 1$. 
As $\kg^2$ represents the ratio of field intensity and frequency, it follows that one 
needs very strong fields at very low frequencies $(2\pi \nu = ck)$: 
\be 
\kg^2 = \lh \frac{0.3 \times 10^{-18} E_0\, (\mbox{V/m})}{2\pi \nu\, (\mbox{Hz})} 
 \rh^2 \geq 1.
\label{2.5.4}
\ee 
An upper limit to the fields strength of the order of $10^{19}$ V/m arises because 
of pair creation processes which destroy the electric field at higher intenties 
\ct{schwinger}. Therefore we obtain an upper limit of the order of $\sim 1$ Hz for the 
frequencies which can give rise to parametric resonance. For lower fields and higher 
frequencies one finds standard periodic behavior described by solutions with $\sg = 0$ 
or imaginary.\nl

\nit
An important characteristic of the motion of a chargeless particle in the background of 
the fields (\ref{1},\ref{6},\ref{7}) is that a particle initially at the origin in the 
transverse plane before the arrival of the wave, remains at the origin at all later 
times. This allows one to translate directly the geodesic motion of a test particle in 
terms of the relative rate of acceleration of this particle w.r.t.\  another one located 
in the origin: {\boldmath{$\xi$}} = 0.  
\nl

\nit
For charged particles this situation changes. If $q \neq 0$ and we take the 
transverse co-ordinate $\xi = \xi_{\|}$ parallel to the electric field, the equation 
of transverse motion is:
\be
\dsp{ \frac{d^2\xi}{du^2} + \frac{\kg^2 k^2}{2}\, \lh 1 + \cos 2ku \rh \xi 
 = - \ve \cos ku, }
\label{2.7}
\ee
with 
\be 
\ve = \frac{qka}{m\gam}\,  = \frac{qE_0}{mc\gam}. 
\label{2.7.1}
\ee 
In the inhomogeneous case the absence of an expontential term in the driving force  
implies that any special solution of the inhomogeneous equation can have an expansion 
(\ref{2.1}) only with $\sg = 0$; for these special solutions the inhomogeneous Mathieu 
equation becomes 
\be 
\lh \ba{cccc} a_0 &  B  &  0  & ... \\
               0  & a_1 &  B  & ... \\
               0  &  0  & a_2 & ... \\
               .  &  .  &  .  &  .   \ea \rh_{\sg = 0}\, \lh \ba{c} \bar{\rg}_0 \\
                                                          \bar{\rg}_1 \\
                                                          \bar{\rg}_2 \\
              . \\ \ea \rh\, = \lh \ba{c} \thg \\ 0 \\ 0 \\ . \ea \rh, 
\hs{2} \thg = \frac{\ve}{k^2}\, \left[ \ba{c} 1 \\ 0 \ea \right]. 
\label{2.7.2}
\ee 
Here we are interested especially in obtaining a particular solution of the previous 
type: $\bar{\rg}_0 \neq 0$, $\bar{\rg}_n = 0$,  $n \geq 1$ . The non-zero component 
$\bar{\rg}_0$ is a solution of the inhomogeneous equation 
\be
a_0 \bar{\rg}_0 = \thg. 
\label{2.8}
\ee 
Now $a_0$ is of the form (\ref{2.4.0}), with $\sg = 0$:
\be 
a_0 = \lh \ba{cc} - 1 + \frac{3\kg^2}{4} + \Del_{11}(0) \kg^4  & 0 \\
              0  & -1 + \frac{\kg^2}{4}\, + \Del_{11}(0) \kg^4 \ea \rh
 + {\cal O}{(\kg^6)},
\label{2.8.1}
\ee 
where beyond the lowest-order approximation we find $\Del_{11}(\sg = 0) = 
\frac{1}{144}$. As a result a special solution of the inhomogeneous equation 
is given by 
\be
\bar{\rg}_0 = \frac{- \ve}{k^2 (1 - \frac{3\kg^2}{4} - \Del_{11}(0) \kg^4 + ...)}\, 
 \left[ \ba{c} 1 \\ 0 \ea \right], \hs{2} \bar{\rg}_n = 0, \hs{1} n\geq 1. 
\label{2.9.4}
\ee 
The general solution of this system is the sum of a particular solution plus a solution 
of the homogeneous equation, which can have $\sg \neq 0$. An important difference with 
the previous case is that a particle initially at the origin does not remain at the 
origin, but is accelerated by the electric field. Thus the study of the relative 
acceleration between particles on different world lines becomes more complicated; for 
a general discussion we refer to  ref.\ \ct{vhrk}.

\section{QED corrections}

As the magnitude of the electric field in the region of parametric resonance is 
necessarily extremely high, QED effects such as vacuum polarization can in principle 
not be ignored. However, as we now show they do not affect the classical wave solutions 
we have studied in the previous paragraphs. 

The non-linear effects of quantum fluctuations on the propagation of light in flat 
(Minkowski) space-time were first studied by Euler and Heisenberg \ct{EH}. The effects 
are summerized by a well-known effective lagrangean incorporating the corrections due 
to the electromagnetic higher-order terms; its generally covariant form reads:
\be
\dsp{\mathscr{L}= \ve_0 c^2\sqrt{-g}\left(-\frac{1}{4}F^2 + 
 \bg \left[ (F^2)^2 + \frac{7}{4}\phi^2 \right] \right)}
\label{3.4}
\ee
where:
\be
\ba{l}
\dsp{F^2 = g^{\mu\kg} g^{\nu\lb} F_{\mu\nu} F_{\kg\lb},}\\
\\
\dsp{~~\phi=F_{\mu\nu}{\tilde{F}}^{\mu\nu},}
\ea
\label{3.5}
\ee
and ${\tilde{F}}_{\mu\nu}$ is the dual electromagnetic tensor defined by:
\be
\tilde{F}^{\mu\nu}=\frac{1}{2\sqrt{-g}}\, \varepsilon^{\mu\nu\rho\sg}F_{\rho\sg}.
\label{3.6}
\ee 
The coupling constant $\bg$  is given by 
\be 
\bg = \frac{\ag^2 \hbar^3 \ve_0}{90 m^4 c^4}. 
\label{3.6.1}
\ee 
We first derive the Maxwell equations from the Euler-Lagrange equations taking $A_{\nu}$ 
and $ \partial_{\nu}{A_{\mu}}$ as the variables: 
\be
\frac{\partial{\mathscr{L}}}{\partial{A_{\nu}}}-\partial_{\mu}{
 \frac{\partial{\mathscr{L}}}{\partial{(\partial_{\mu}{A_{\nu}})}}}=0.
\label{3.8}
\ee
Combining (\ref{3.4}) and (\ref{3.8}), we get the following equations:
\be
\ba{l}
\dsp{\nabla_{\mu}\left[ (1 - 8\bg F^2) F^{\mu\nu} 
 - 14 \bg \phi \tilde{F}^{\mu\nu}\right] = 0. }
\ea
\label{3.9}
\ee
The energy-momentum tensor is obtained from the lagrangean (\ref{3.4}) by:
\be
\ba{rcl}
\dsp{T_{\mu\nu} }& \hs{-.5} = & \dsp{ \hs{-.5} 
 \frac{-2}{\sqrt{-g}} \frac{ \partial{\mathscr{L}} }{ \partial{g^{\mu\nu}} } }\\
 & & \\
 &  \hs{-.5} = & \dsp{ \hs{-.5} 
 \ve_0 c^2 \left[(1 - 8\bg F^{2}) F_{\mu\lambda}{F_{\nu}}^{\lambda} -
\frac{1}{4} g_{\mu\nu} ( F^2  - 4 \bg (F^2)^2) - \frac{7\bg}{4}\,  g_{\mu\nu}\phi^2 
 \right]. }
\ea
\label{4.1}
\ee
With the metric (\ref{1})  the Ricci tensor $R_{\mu \nu}$ has only one non-zero 
component $R_{uu}$, the Einstein field  equations reduce to:
\be
\dsp{R_{uu}=-\frac{1}{2}\Delta_{trans}K=8 \pi G T_{uu}}, \hs{2} 
 \mbox{all other}\,  T_{\mu\nu} = 0.
\label{4.2}
\ee
In our particular case of a plane-wave vector potential of the form (\ref{6}), the 
only non-vanishing components of the electromagnetic tensor $F_{\mu \nu}$ are:
\be
\dsp{F_{ui}=-F_{iu}=a_{i}k\cos{ku},  \hs{2} i=(x,y). }
\label{4.3}
\ee
It follows that $F^2 = \phi = 0$, and the Maxwell-Einstein field equations 
reduce to the original form:
\be
\ba{l}
\dsp{\nabla_{\mu} F^{\mu\nu}=0, }\\
\\
\dsp{T_{uu}= - \ve_0 c^2\, F_{u\lambda}{F_{u}}^{\lambda}. }
\ea
\label{4.4}
\ee
Thus the quantum corrections do not modify the gravito-electro-magnetic plane wave 
solution. 


\np
\nit
{\bf Appendix} \nl

\nit
The solutions of the Mathieu equation are classified as having an odd or even 
periodic part. Eq.(\ref{2.1}) represents the general solution of odd type. The general 
solution of even type can be expanded as 
\be 
\xi(u) = e^{\sg ku}\,  \lh \frac{\mu_0}{2}\, + \sum_{n=1}^{\infty} \left[ \mu_n 
 \cos 2nku + \nu_n \sin 2nku \right] \rh.
\label{a.1}
\ee 
After multiplication with $e^{-\sg ku}$ the homogeneous Mathieu equation becomes 
\be
\ba{lll}
0 & = & \dsp{ \lh \frac{1}{2}\, \sg^2 + \frac{1}{4}\, \kg^2 \rh \mu_0\, 
  + \frac{\kg^2}{4}\, \mu_1  }\\ 
 & & \\
 & & \dsp{ +\, 
 \sum_{n= 1}^{\infty} \left[ \lh \sg^2 - 4 n^2 + 
 \frac{\kg^2}{2} \rh \mu_n + 4 n\sg \nu_n   + \frac{\kg^2}{4}\, \lh \mu_{n-1}
 + \mu_{n+1} \rh\right] \cos 2nku }\\
 & & \\
 & & \dsp{ +\,  \sum_{n=1}^{\infty} \left[ \lh \sg^2 - 4 n^2 + 
 \frac{\kg^2}{2} \rh \nu_n - 4 n\sg \mu_n \right] \sin 2nku }\\
 & & \\ 
 & & \dsp{ +\, \frac{\kg^2}{4}\, \lh \nu_2 \sin 2ku + 
 \sum_{n=2}^{\infty} (\nu_{n-1} + \nu_{n+1}) \sin 2nku \rh . }
\ea 
\label{a.2}
\ee 
It follows, that
\be
\mu_0 = - \frac{\kg^2\mu_1}{2\sg^2 + \kg^2},
\label{a.3}
\ee
and 
\be 
\lh \ba{cccc} D_1 &  B  &  0  & ... \\
               B  & D_2 &  B  & ... \\
               0  &  B  & D_3 & ... \\
               .  &  .  &  .  &  .  \ea \rh \lh \ba{c} \zg_1 \\ 
                                                       \zg_1 \\
                                                       \zg_2 \\ 
                                                        . \ea \rh\, = 0, 
\label{a.4}
\ee 
where $B$ is the $2 \times 2$-matrix (\ref{2.3.1}), whilst 
\be 
\ba{l} 
\dsp{ D_1 = \lh \ba{cc} 
 \sg^2 - 4 + \frac{\kg^2}{2} - \frac{\kg^4}{4 (2\sg^2 + \kg^2)} & 4 \sg \\
  - 4 \sg & \sg^2 - 4 + \frac{\kg^2}{2} \ea \rh, } \\ 
 \\
\dsp{ D_n =  \lh \ba{cc} 
 \sg^2 - 4n^2 + \frac{\kg^2}{2} & 4 n \sg \\
  - 4 n\sg & \sg^2 - 4n^2 + \frac{\kg^2}{2} \ea \rh, } \hs{3} n \geq 2, 
\ea 
\label{a.4.1}
\ee 
and 
\be
\zg_n = \lh \ba{c} \mu_n \\ \nu_n \ea \rh.
\label{a.5}
\ee
To solve this equation, we bring the infinite-dimensional matrix equation (\ref{a.4}) 
into upper triangular form: 
\be 
\lh \ba{cccc} D_1 &  B  &  0  & ... \\
               B  & D_2 &  B  & ... \\
               0  &  B  & D_3 & ... \\
               .  &  .  &  .  &  .  \ea \rh = \lh\ba{cccc} d_1 &  B  &  0  & ... \\
               0  & d_2 &  B  & ... \\
               0  &  0  & d_3 & ... \\
               .  &  .  &  .  &  .  \ea \rh \lh \ba{cccc} 1 &  0  &  0  & ... \\
               c_1 & 1 &  0  & ... \\
               0  &  c_2  & 1 & ... \\
               .  &  .  &  .  &  .  \ea \rh, 
\label{a.6}
\ee 
with $d_n$ and $c_n$ the solutions of the equations 
\be 
d_n + B c_n = D_n, \hs{2} d_{n+1} c_n = B \hs{1} \Rightarrow \hs{1} 
 B^{-1} d_n + (B^{-1} d_{n+1})^{-1}  = B^{-1} D_n. 
\label{a.7}
\ee 
The solution for $d_n$ takes the form of a continuing fraction 
\be 
d_n = D_n - B^2 \left[ D_{n+1} - B^2 \left[ D_{n+2} - ... \right]^{-1} \right]^{-1},
\label{a.8}
\ee 
where we have used the fact that $B$ is proportional to the unit matrix, hence
commutes with all other matrices. If we define
\be 
\lh \ba{c} \bar{\zg}_1 \\ \bar{\zg}_2 \\ \bar{\zg}_3 \\ . \ea \rh = 
 \lh \ba{cccc} 1 &  0  &  0  & ... \\
               c_1 & 1 &  0  & ... \\
               0  &  c_2  & 1 & ... \\
           .  &  .  &  .  &  .  \ea \rh \lh \ba{c} \zg_1 \\ \zg_2 \\ \zg_3 \\ . \ea \rh,
\label{a.9}
\ee 
equivalent with $\bar{\zg}_1 = \zg_1$, and
\be
\bar{\zg}_n = \zg_n + c_{n-1} \zg_{n-1}, \hs{2} n\geq 2.
\label{a.10}
\ee 
Now the equation 
\be 
\lh\ba{cccc} d_1 &  B  &  0  & ... \\
               0  & d_2 &  B  & ... \\
               0  &  0  & d_3 & ... \\
               .  &  .  &  .  &  .  \ea \rh 
               \lh \ba{c} \bar{\zg}_1 \\ \bar{\zg}_2 \\ \bar{\zg}_3 \\ . \ea \rh = 0, 
\label{a.11}
\ee 
has solutions with $\bar{\zg}_N \neq 0$, and 
\be 
d_k \bar{\zg}_k = - B \bar{\zg}_{k+1}, \hs{1} 1 \leq k \leq N-1 \hs{1} \mbox{and} \hs{1}  
\bar{\zg}_n = 0, \hs{1} n > N.
\label{a.12}
\ee 
The first one of these is the one given by $d_1 \bar{\zg}_1 = 0$, $\bar{\zg}_n = 0$, 
$n \geq 2$.
\nl

\nit 
{\em Approximations.\/} Solutions of the equation $d_1 \bar{\zg}_1 = 0$ exist if 
$\det d_1 = 0$. To solve this condition, we need to compute the infinite fraction 
(\ref{a.8}). We can do this by successive approximations: 
\be 
\ba{lll}
d^{(0)}_1 & = & D_1, \\
 & & \\
d_1^{(1)} & = & \dsp{ D_1 - B^2 D^{-1}_2 = \lh D_1 D_2 - B^2 \rh D_2^{-1}, }\\
 & & \\ 
d_1^{(2)} & = & \dsp{ \lh D_1 D_2 D_3 - B^2 D_1 - B^2 D_3 \rh 
 \lh D_2 D_3 - B^2 \rh^{-1}, }\\
 & & \\
 ... 
\ea 
\label{a.13}
\ee 
To zeroth order approximation, we can show that $\det D_1$ = 0 has solutions for
$8 < \kg^2 < 16$: 
\be
\left| D_1 \right| = \frac{\kg^2}{8}\, \lh \kg^2 - 8 \rh\lh \kg^2 - 16 \rh 
 + 2 \sg^2 \lh \sg^4 + \lh 8 + \frac{3\kg^2}{2} \rh \sg^2 + 16 + \frac{5\kg^2}{8} \rh.
\label{a.14}
\ee 
It suffices to note, that the first term is negative in the range $\kg^2 \in (8,16)$, 
whilst for any $\kg^2$ the second term can take all  positive values between 
$(0, \infty)$.

\end{document}